\documentclass[graybox]{svmult}

\newcommand{\aFe}{\ensuremath{\alpha/{\rm Fe}}}

\usepackage{mathptmx}       % selects Times Roman as basic font
\usepackage{helvet}         % selects Helvetica as sans-serif font
\usepackage{courier}        % selects Courier as typewriter font
\usepackage{type1cm}        % activate if the above 3 fonts are
\usepackage{makeidx}         % allows index generation
\usepackage{graphicx}        % standard LaTeX graphics tool
\usepackage{multicol}        % used for the two-column index
\usepackage[bottom]{footmisc}% places footnotes at page bottom

\makeindex             % used for the subject index
\begin{document}

\title*{And the winner is: galaxy mass}
\author{Daniel Thomas}
\institute{Daniel Thomas \at Institute of Cosmology and Gravitation, University of Portsmouth, Dennis Sciama Building, Burnaby Road, Portsmouth, PO1 3FX, UK, \email{daniel.thomas@port.ac.uk}}

\maketitle

% Too much empty space in the original style file!
\vskip-1.2truein

\abstract{The environment is known to affect the formation and evolution of galaxies considerably best visible through the well-known morphology-density relationship. We study the effect of environment on the evolution of early-type galaxies for a sample of 3,360 galaxies morphologically selected by visual inspection from the SDSS in the redshift range $0.05\leq z\leq 0.06$, and analyse luminosity-weighted age, metallicity, and \aFe\ ratio as function of environment and galaxy mass. We find that on average 10 per cent of early-type galaxies are rejuvenated through minor recent star formation. This fraction increases with both decreasing galaxy mass and decreasing environmental density. However, the bulk of the population obeys a well-defined scaling of age, metallicity, and \aFe\ ratio with galaxy mass that is independent of environment. Our results contribute to the growing evidence in the recent literature that galaxy mass is the major driver of galaxy formation. Even the morphology-density relationship may actually be mass-driven, as the consequence of an environment dependent characteristic galaxy mass coupled with the fact that late-type galaxy morphologies are more prevalent in low-mass galaxies.}

\section{Introduction}
\label{sec:intro}
The environment is known to be a major driver in the formation and evolution of galaxies. Its influence is best visible through the well-known morphology-density relationship, according to which
early-type galaxies and morphologically undisturbed galaxies are
preferentially found in high density environments and vice versa
\cite{Dressler80,PG84}. 

In contrast, it is less clear whether the environment is equally
important at a given galaxy morphology. There is still major
controversy about whether the formation and evolution of the most
massive and morphologically most regular galaxies in the universe,
i.e.\ early-type galaxies, are affected by environmental
densities. We analyse the stellar population parameters luminosity-weighted age, metallicity, and \aFe\ element ratio of 3,360 early-type galaxies drawn from the Sloan Digital Sky Survey \cite{Yorketal00} in a narrow redshift range ($0.05\leq z\leq 0.06$) \cite{Thomas10a}. 
The SDSS provides the opportunity to explore huge homogeneous samples of early-type galaxies in the nearby universe, comprising several ten thousands of objects, so that a statistically meaningful investigation of the stellar population parameters of galaxies and their dependence on environment can be attempted. This work is published in Thomas et al.\ (2010), and we refer the reader to this paper for more details \cite{Thomas10a}. The galaxy catalogue produced in this study can be found at www.icg.port.ac.uk/$\sim$thomasd.

\section{Data}
\label{sec:data}
The sample utilised here is part of a project called MOSES:
\textbf{MO}rphologically \textbf{S}elected \textbf{E}arly-types in
\textbf{S}DSS. We have collected a magnitude limited sample of 48,023 galaxies in the redshift range $0.05\leq z\leq 0.1$ with apparent $r$-band magnitude brighter than 16.8 from the SDSS Data Release 4. The most radical difference with respect to other galaxy samples constructed from SDSS is our choice of {\rm purely morphological} selection of galaxy type through visual inspection \cite{Schawetal07}. Building on the success of this strategy, our approach has been extended with the Citizen Science Galaxy Zoo project to enable visual classification of even larger samples \cite{Lintott08}.

For the estimate of the environmental density we calculate the local number density of objects brighter
than a certain absolute magnitude in a $6\;$Mpc sphere around the object of
interest with a Gaussian weight function centred on the object ensures to avoid contamination from neighbouring structures. We compute stellar velocity dispersion and measure emission line fluxes, which we use to produce a emission line free galaxy spectrum with the public codes ppxf and GANDALF \cite{CE04,Sarzietal06}. We measure Lick absorption-line indices on the spectra, and derive luminosity-weighted ages, metallicities, and \aFe\ ratios using the TMB stellar population models \cite{TMB03a} by means of a minimised $\chi^2$ technique. 

We applied further selection criteria to the 16,502 early-type galaxies in our catalogue. The redshift range sampled ($0.05\leq z\leq 0.1$) is small and corresponds to a time window of only $\sim 600\;$Myr. Still, this age difference produces significant selection effects. We therefore took a conservative approach and decided to focus on the narrow redshift range $0.05\leq z\leq 0.06$, providing an acceptable coverage in velocity dispersion down to $\log\sigma/{\rm km/s}\sim 1.9$. This results in a total of 3,360 early-type galaxies.

\section{Results}
\label{sec:results}
The characteristic mass of a galaxy population increases
with environmental density \cite{Baldryetal06}, indicating that lower density
environments host less massive galaxies. This mass bias has to be
eliminated for a meaningful study of the influence of environment. We
therefore investigate the environmental dependence of the
correlations of the stellar population properties with velocity
dispersion $\sigma$ and galaxy mass.

\begin{figure}[ht]
\includegraphics[width=0.49\textwidth]{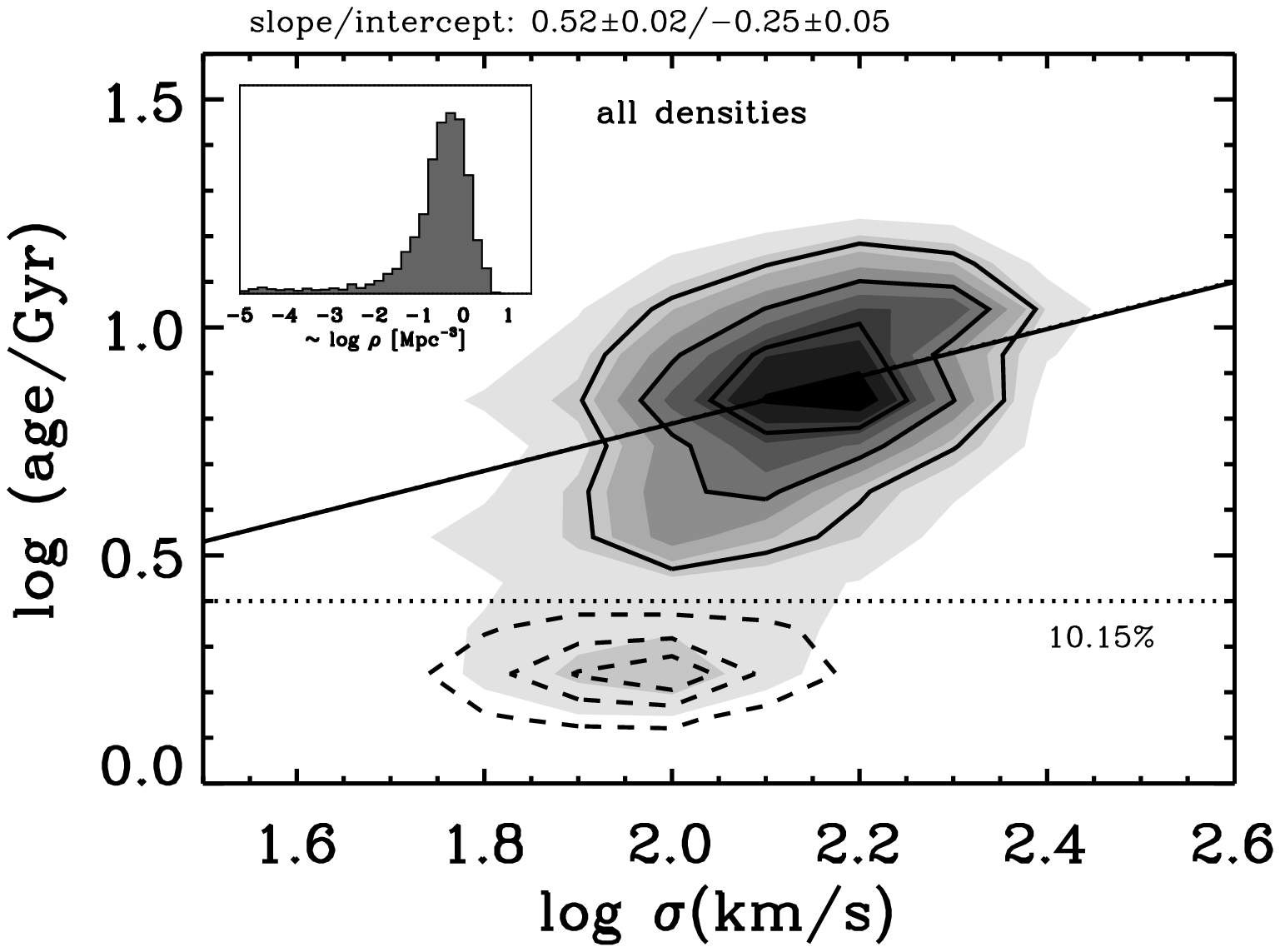}
\includegraphics[width=0.49\textwidth]{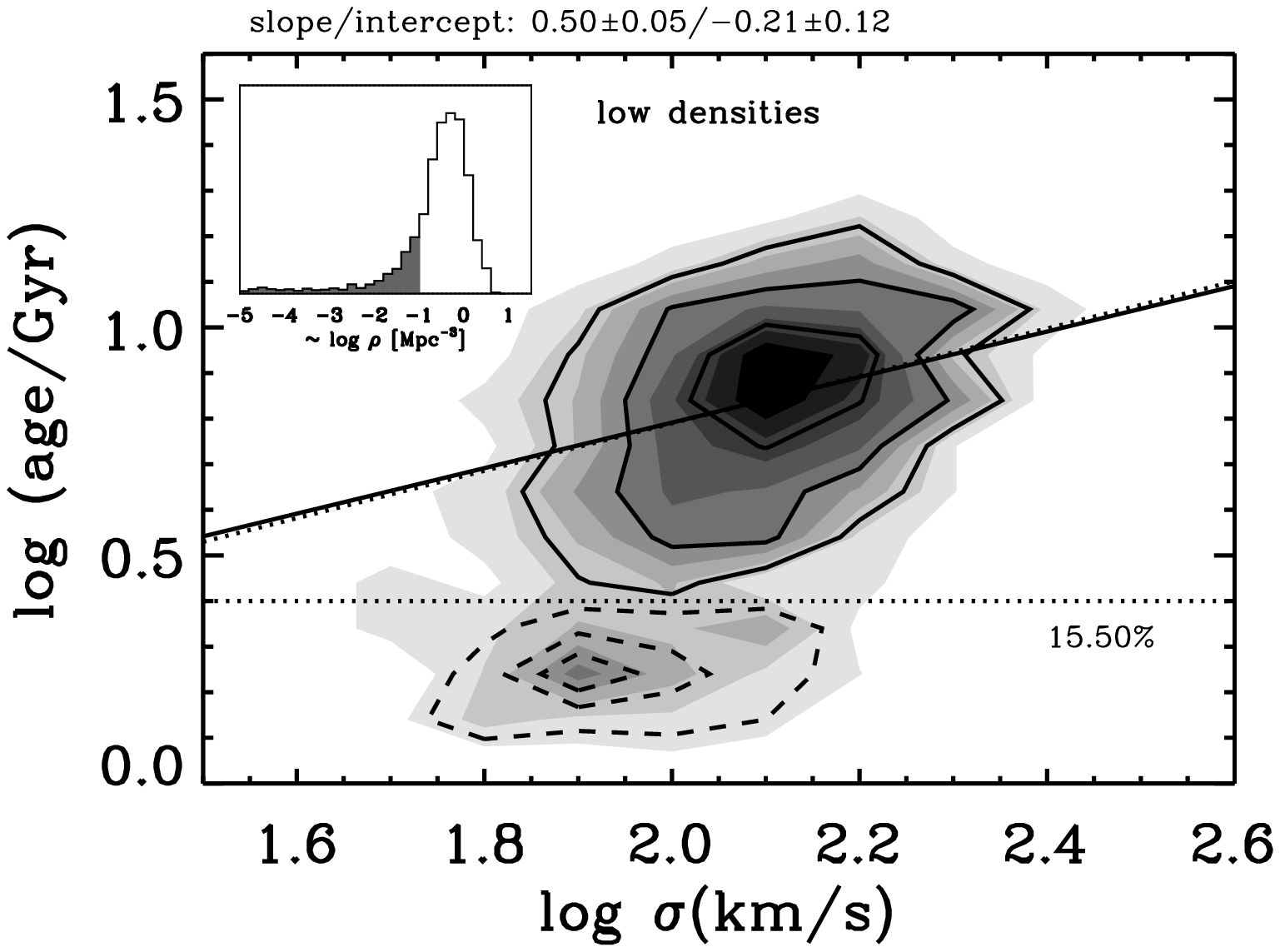}
\includegraphics[width=0.49\textwidth]{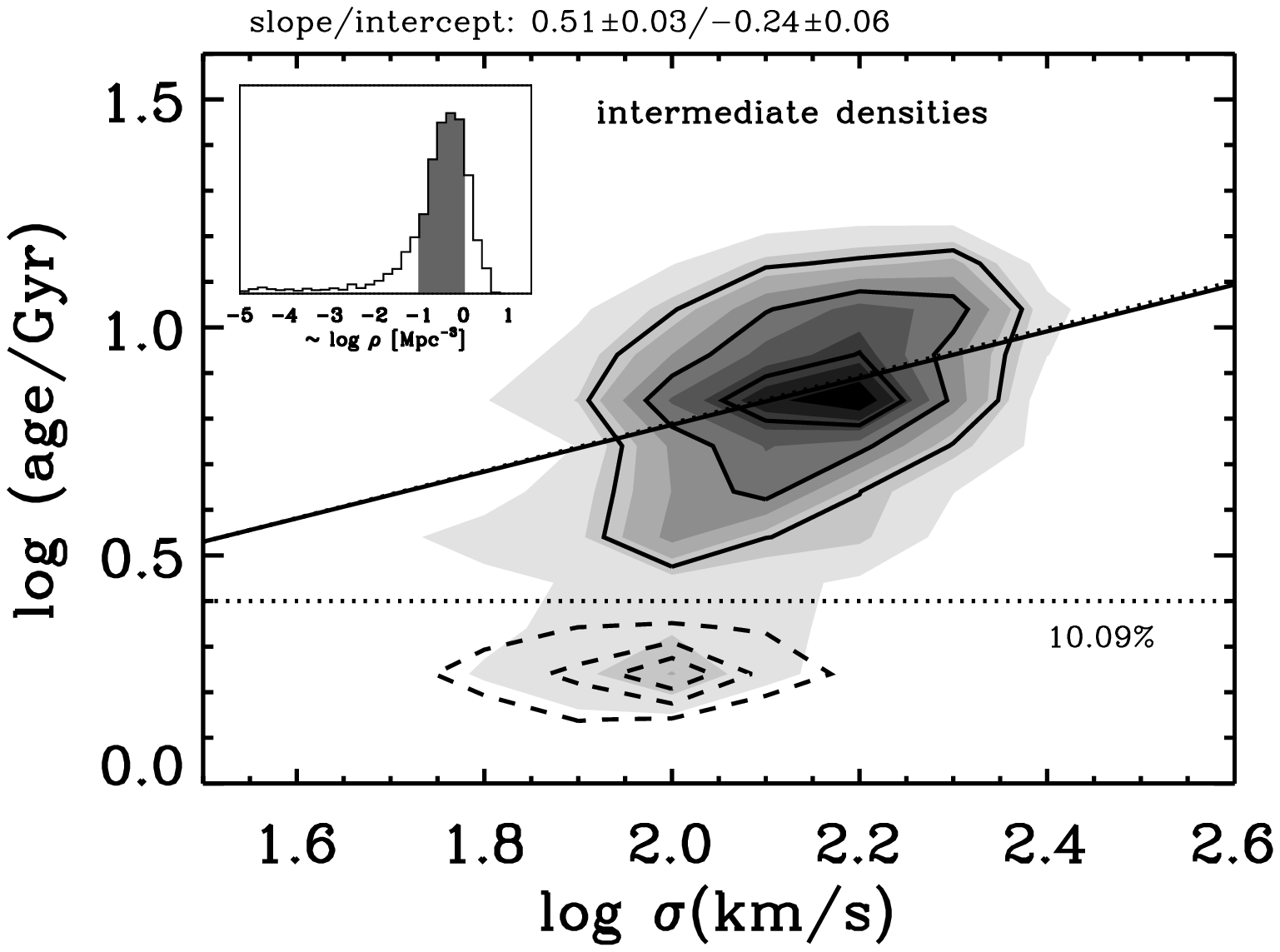}
\includegraphics[width=0.49\textwidth]{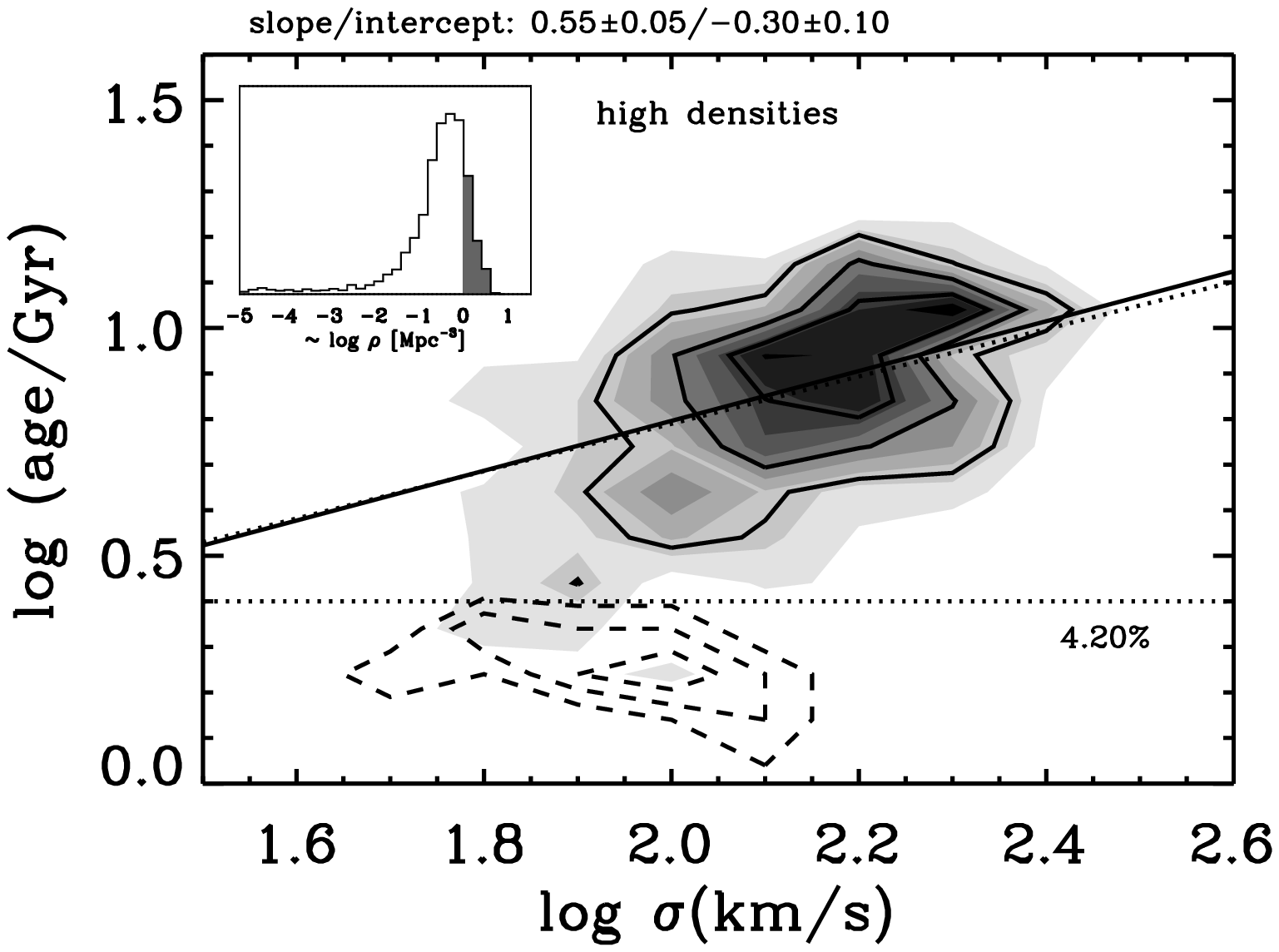}
\caption{Contour plots of the relationship between stellar velocity dispersion and luminosity-weighted age for various environmental densities as indicated by the inset histograms. The horizontal dotted line separates an old red sequence population (solid contours) from rejuvenated objects in the blue cloud with light-averaged ages smaller than $2.5\;$Gyr (dashed contours). The solid line is a linear fit to the red sequence population, the parameters of the fit are given at the top of each panel. The dotted line is the fit for all environmental densities (parameters from top left-hand panel). The age-$\sigma$ relationship for the red sequence population is independent of environment, while the rejuvenation fraction increases with decreasing density.}
\label{fig:agestat}
\end{figure}

Fig.~\ref{fig:agestat} shows contour plots for the relationship between stellar velocity dispersion and
luminosity-weighted age for various environmental densities as indicated by the inset histograms. The distribution of ages is bimodal with a major peak at old ages and a secondary peak at young ages around $\sim 2.5\;$Gyr in analogy to 'red sequence' and 'blue cloud' identified in galaxy populations. We call the objects below the dotted line 'rejuvenated', and the fraction of objects in this category 'rejuvenation fraction'. The latter is about 10 per cent as indicated by the label in the figure.

The fits to the red sequence population are repeated in every environment bin. It can be seen clearly from the plots that the resulting fit parameters are consistent with no variation as a function of environment within their 1-$\sigma$ error bars. We conclude there is no considerable change as a function of environmental density, hence the age-$\sigma$ relationship is independent of environment. The same is true for metallicity and \aFe\ ratio \cite{Thomas10a}.

Very different is the behaviour of the blue cloud (rejuvenated) population. In this case, the environment plays a role. The rejuvenation fraction increases with decreasing environmental density. Hence early-type galaxies in lower density environments are not generally younger (at a given mass), but the fraction of rejuvenated galaxies is higher. Note, however, that again mass is the major driver for the fraction of rejuvenated galaxies as shown in Thomas et al (2010) \cite{Thomas10a}.

This environment-dependent rejuvenation process is most prevalent in low-mass galaxies, and entirely absent in the most massive objects. Most importantly, rejuvenation involves only minor star formation, hence is negligible in terms of the overall mass budget, and occurs at late epochs. As a consequence, most of the galaxy's stellar populations form in an environment independent mode with galaxy mass as the major driver for star formation and quenching processes.

\section{Discussion}
In Thomas et al (2010) we investigate the effect of environmental density on the formation epochs of early-type galaxies \cite{Thomas10a}. The major conclusion is that it is galaxy mass, not environment, that shapes the stellar population properties of early-type galaxies. Pre-SDSS studies of the stellar population parameters in early-type galaxies based on relatively small, local samples consistently found younger average ages in low density environments \cite{Kuntschner02,Thomas05}. The present sample based on SDSS is significantly larger and therefore less biased towards the local volume, which may explain the discrepancy with previous work. The increase in sample size further implies a higher statistical significance of the results. In fact most studies in the recent literature that are based on large samples from data bases such as the SDSS come to similar conclusions \cite{vandenBosch08,Peng10,Gruetzbauch10,Rogers10}: galaxy mass dominates over galaxy environment. In particular, environment seems to be important mostly for low-mass galaxies and only at late epochs \cite{Tasca09,Pasquali10}.

Based on a study of SDSS data, van den Bosch (2008) show that the colours and concentrations of satellite galaxies are determined by their stellar mass. In particular, at fixed stellar mass, they find the average colours and concentrations of satellite galaxies to be independent of either halo mass or
halo-centric radius \cite{vandenBosch08}. The only fingerprint from the environment is that satellites appear to be redder, older and metal-richer than centrals of the same stellar mass, a difference that increases with decreasing galaxy mass \cite{Pasquali10}. However, they find that the nature of the transformation and quenching process experienced by a galaxy when it falls into a bigger halo is independent of the size of this halo, hence independent of environment \cite{vandenBosch08}.

Peng et al (2010) use SDSS data to separate two distinct processes of "mass-quenching" and "environment-quenching" that dominate the evolution of galaxies \cite{Peng10}. They find that the quenching of galaxies around and above $M^{*}$ must follow a rate that is statistically proportional to their star-formation rates, and the latter tend to scale with mass rather than with environment. Peng et al (2010) conclude that the environment acts through a "once-only" process as the environment of a given galaxy changes, while the mass-quenching process must be continuously operating and be governed by a probabilistic transformation rate \cite{Peng10}. Again, this result implies that environment only plays a secondary role in shaping the stellar populations of galaxies, and galaxy mass is the major driver of galaxy formation.

\section{Conclusions}
The role and significance of the environment for galaxy formation has been discussed highly controversially at the JENAM 2010 Symposium 2 'Environment and the Formation of Galaxies: 30 years later'.  It is exciting and intriguing that recent work is about to lead to a converging picture, despite some controversy, in which environment loses out over galaxy mass, 30 years after the seminal work of Alan Dressler on the morphology-density relationship \cite{Dressler80}. As it turned out at this meeting, however, Alan Dressler has never regarded the morphology-density relation necessarily as environment-driven. It could actually be mass-driven, being the consequence of an environment dependent galaxy mass function (the characteristic mass of a galaxy population increases
with environmental density) coupled with the fact that late-type galaxy morphologies are more prevalent in low-mass galaxies. Galaxies are individualists, and, as Alan Dressler concluded at the meeting, 'morphology is destiny'.

%%% Bibliography

%\input{referenc}
\end{document}